\documentclass[preprint,showpacs,preprintnumbers,amsmath,amssymb]{revtex4}

\def\undersim#1{\setbox9\hbox{${#1}$}{#1}\kern-\wd9\lower
    2.5pt \hbox{\lower\dp9\hbox to \wd9{\hss $_\sim$\hss}}}
\usepackage{amssymb}
\usepackage{amsmath}
\usepackage{graphicx}
\usepackage{color}

\def\undersim#1{\setbox9\hbox{${#1}$}{#1}\kern-\wd9\lower
    2.5pt \hbox{\lower\dp9\hbox to \wd9{\hss $_\sim$\hss}}}

\def\mk{{\mathbf k}}

\begin{document}

\title{Squeezed back-to-back correlations of $\phi\phi$ in Au$+$Au
and d$+$Au collisions at the Relativistic Heavy Ion Collider energies}

\author{Yong Zhang$^1$}
\author{Wei-Ning Zhang$^{1,\,2,}$\footnote{wnzhang@dlut.edu.cn}}
\affiliation{$^1$School of Physics, Dalian University of Technology,
Dalian, Liaoning 116024, China\\
$^2$Department of Physics, Harbin Institute of Technology, Harbin,
Heilongjiang 150006, China}


\begin{abstract}
We investigate the squeezed back-to-back correlations (BBC) of $\phi\phi$,
caused by the mass modification of the particles in the source medium,
in the heavy-ion collisions of Au+Au and d+Au at the Relativistic Heavy
Ion Collider (RHIC) energies.  The BBC functions are calculated using the
modified masses extracted from experimental data and the source space-time
distributions provided by the viscous hydrodynamic code VISH2+1.
Our investigations indicate that the BBC of $\phi\phi$ may perhaps be
observed in the collisions of d+Au and the peripheral collisions of Au+Au
at the RHIC.  We suggest to measure the BBC experimentally for understanding
the mass modifications of the $\phi$ meson in the collisions. \\[2ex]
\end{abstract}


\maketitle

The main goal of high-energy heavy-ion collisions is to create the matter
of quark-gluon plasma (QGP), predicted by quantum chromodynamics, and study
its properties \cite{BRAHMS-NPA05,PHOBOS-NPA05,STAR-NPA05,PHENIX-NPA05}.
As there are copious strange quarks $s$ and $\bar s$ in the QGP environment,
$\phi(s\bar s)$ meson can produce readily, bypassing the Okubo-Zweig-Izuka
(OZI) rules \cite{OZI}.  The productions of the $\phi$ meson observed
in the heavy-ion collisions at the Relativistic Heavy Ion Collider (RHIC)
and the Large Hadron Collider (LHC) give a certain evidence of having formed
the QGP matter in the experiments \cite{{STAR-PRC02,STAR-PLB05,PHENIX-PRC05,
STAR-PLB09,STAR-PRC09,STAR-PRL07,MohantyXu-JPG09,ALICE-PRC15}}.
On the other hand, the weak interactions expected between the $\phi$ meson
and the hadronic mediums makes it as a sensitive probe of the QGP properties
\cite{{STAR-PRL07,STAR-PRC09,MohantyXu-JPG09,Shor-PRL85,ChenMa-PRC06,
PHENIX-PRL07,Hirano-PRC08,Nasim-PRC13,Takkeuchi-PRC15,STAR-PRL16}}.
The investigations of $\phi$ elliptic flow in the heavy-ion collisions at
the RHIC energies indicate that the flow reflects dominantly the anisotropy
of the QGP matter and the effect of hadronic scattering is unimportant
\cite{{STAR-PRL07,STAR-PRC09,MohantyXu-JPG09,ChenMa-PRC06,PHENIX-PRL07,
Hirano-PRC08,Nasim-PRC13,Takkeuchi-PRC15,STAR-PRL16}}.  However, the $\phi$
meson is also argued to be with a larger hadronic cross section
than the estimations by current theories, based on the recent measurements
of the elliptic flow of identified hadrons in the Pb-Pb collisions at
$\sqrt{s_{NN}}=2.76$ GeV at the LHC \cite{ALICE-JHEP15}.  It is still an
open issue to determine the interaction between $\phi$ and the hadronic
medium.

In the particle-emitting sources formed in high-energy heavy-ion collisions,
the interaction between the particle and source medium may lead to a
modification of the particle mass, and thus give rise to a squeezed
correlation of boson-antiboson \cite{AsaCso96,AsaCsoGyu99}.  This squeezed
correlation is the result of a quantum mechanical transformation relating
in-medium quasiparticles to the two-mode squeezed states of their free
observable counterparts, through a Bogoliubov transformation between the
creation (annihilation) operators of the quasiparticles and the free
observable particles \cite{AsaCso96,AsaCsoGyu99}.  Because the correlation
impels the boson and antiboson to move in opposite directions, it is known
as back-to-back correlation (BBC) \cite{AsaCso96,AsaCsoGyu99,Padula06}.
The measurements of the BBC of $\phi\phi$ pair may give a knowledge of
the interaction between the $\phi$ meson and the source medium, and perhaps
provide a new way to probe the thermal properties of the hadronic sources
\cite{AsaCso96,AsaCsoGyu99,Padula06,Padula10,Zhang15a}.

Recently, the BBC of $\phi\phi$ and $K^+K^-$ are calculated with ideal
hydrodynamic sources \cite{Zhang15}.  The calculations indicate that
the BBC of $\phi\phi$ may perhaps be observed in the peripheral heavy-ion
collisions at the RHIC and LHC \cite{Zhang15}, because the temporal
distribution of the source is narrower in the peripheral collisions.
As the BBC is caused by the particle-mass modification in the medium, it
is crucial to use suitable modified masses in examining the BBC.  In this
work, we analyze the mass modification of the $\phi$ in the source medium
with the experimental data measured by the STAR collaboration \cite{STAR-PRC09}.
The BBC of $\phi\phi$ in the Au$+$Au and d$+$Au collisions at the RHIC energies
are investigated, using the viscous hydrodynamic model VISH2$+$1 \cite{VISH2+1}
and the modified masses extracted from the experimental data \cite{STAR-PRC09}.
The investigations indicate that the BBC of $\phi\phi$ may perhaps be observed
in the collisions of d+Au and the peripheral collisions of Au+Au at the RHIC.

The BBC function of the two $\phi$ mesons with momenta $\mk_1$ and $\mk_2$ is
defined as \cite{AsaCsoGyu99,Padula06}
\begin{equation}
\label{BBCf}
C(\mk_1,\mk_2) = 1 + \frac{|G_s(\mk_1,\mk_2)|^2}{G_c(\mk_1,\mk_1) G_c(\mk_2,
\mk_2)},
\end{equation}
where $G_c(\mk_1,\mk_2)$ and $G_s(\mk_1,\mk_2)$ are the chaotic and squeezed
amplitudes, respectively.  For evolution particle-emitting sources, they can
be expressed as
\cite{MakhSiny,AsaCsoGyu99,Padula06,Padula10,Zhang15a}
\begin{equation}
\label{Gchydro}
G_c({\mk_1},{\mk_2})\!=\!\int\! \frac{d^4\sigma_{\mu}(r)}{(2\pi)^3}
K^\mu_{1,2}\, e^{i\,q_{1,2}\cdot r}\,\! \Bigl\{|c'_{\mk'_1,\mk'_2}|^2\,
n'_{\mk'_1,\mk'_2}+\,|s'_{-\mk'_1,-\mk'_2}|^2\,[\,n'_{-\mk'_1,-\mk'_2}+1]\Bigr\},
\end{equation}
\begin{equation}
\label{Gshydro}
G_s({\mk_1},{\mk_2})\!=\!\int\! \frac{d^4\sigma_{\mu}(r)}{(2\pi)^3}
K^\mu_{1,2}\, e^{2 i\,K_{1,2}\cdot r}\!\Bigl\{s'^*_{-\mk'_1,\mk'_2}
c'_{\mk'_2,-\mk'_1}n'_{-\mk'_1,\mk'_2}+c'_{\mk'_1,-\mk'_2} s'^*_{-\mk'_2,\mk'_1}
[n'_{\mk'_1,-\mk'_2} + 1] \Bigr\},
\end{equation}
where $d^4\sigma_{\mu}(r)$ is the four-dimension element of freeze-out
hypersurface, $q^{\mu}_{1,2}=k^{\mu}_1-k^{\mu}_2$, $K^{\mu}_{1,2}=
(k^{\mu}_1+k^{\mu}_2)/2$, and $\mk_i'$ is the local-frame momentum
corresponding to $\mk_i~(i=1,2)$.  In Eqs.\ (\ref{Gchydro}) and
(\ref{Gshydro}), the quantities $c'_{\mk'_1,\mk'_2}$ and $s'_{\mk'_1,
\mk'_2}$ are the coefficients of Bogoliubov transformation between
the creation (annihilation) operators of the quasiparticles and the
free particles, and $n'_{\mk'_1,\mk'_2}$ is the boson distribution
associated with the particle pair
\cite{AsaCso96,AsaCsoGyu99,Padula06,Padula10,Zhang15a}.

\begin{figure}[htbp]
\includegraphics[scale=0.5]{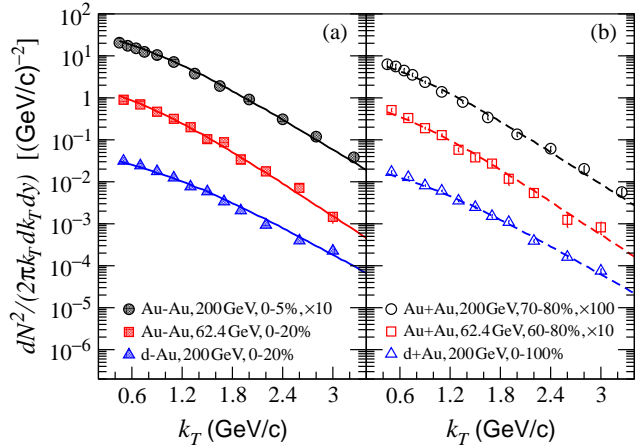}
\caption{Transverse momentum spectra of $\phi$ meson calculated with
VISH2+1 for the Au+Au and d+Au collisions at the RHIC energies.  The
experimental data are from the STAR Collaboration measurements
\cite{STAR-PRC09}. }
\label{zFig_spe}
\end{figure}

We use the VISH2$+$1 code \cite{VISH2+1} to simulate the evolutions of
the particle-emitting sources produced in the collisions of Au$+$Au and
d$+$Au at the RHIC energies.  The event-by-event initial conditions of
MC-Glb \cite{VISHb} are employed at $\tau_0=0.6$ fm/$c$ in the simulations,
and the ratio of the shear viscosity to entropy density of the QGP is
taken to be 0.08 \cite{Shen11-prc,Qian16-prc}.
Figure \ref{zFig_spe}(a) and (b) show the transverse-momentum spectra of
the $\phi$ meson calculated with the viscous hydrodynamic code at the
freeze-out temperature $T_f=140$ MeV for the collisions in different
centrality ranges.  The simulated events for each case is 1000.
Here, the experimental data measured by the STAR collaboration
\cite{STAR-PRC09} are also plotted.  One can see that the calculated
spectra suit the experimental data well.

\begin{figure}[htbp]
\includegraphics[scale=0.60]{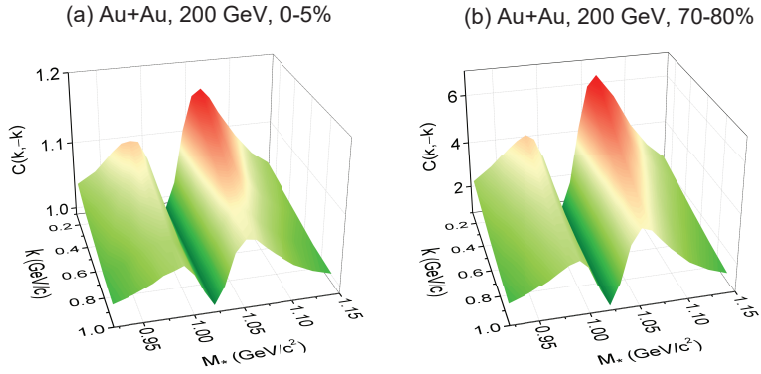}
\vspace*{-8mm}
\caption{BBC functions of $\phi\phi$ for the central and peripheral collisions
of Au+Au at $\sqrt{s_{NN}}=200$ GeV. }
\label{BBC-3D}
\end{figure}

We plot in Fig.\ \ref{BBC-3D} the BBC functions $C(\mk,-\mk)$ in mass-momentum
($M_*$-$k$) plane averaged over the 1000 simulated events of the Au$+$Au
collisions at $\sqrt{s_{NN}}=200$ GeV and in the centrality ranges 0--5\%
and 70--80\%, respectively.
After event average, the oscillations of the BBC functions
with the momentum are smoothed \cite{Zhang15a,Zhang15}.  The BBC function
is sensitive to the mass shift, $M_*-M_0$ ($M_0=1.02$ GeV/$c^2$).  It is
large for the peripheral collisions because the temporal distribution of
the source for the peripheral collisions is narrow \cite{Zhang15}.

\begin{figure}[!htbp]
\includegraphics[scale=0.5]{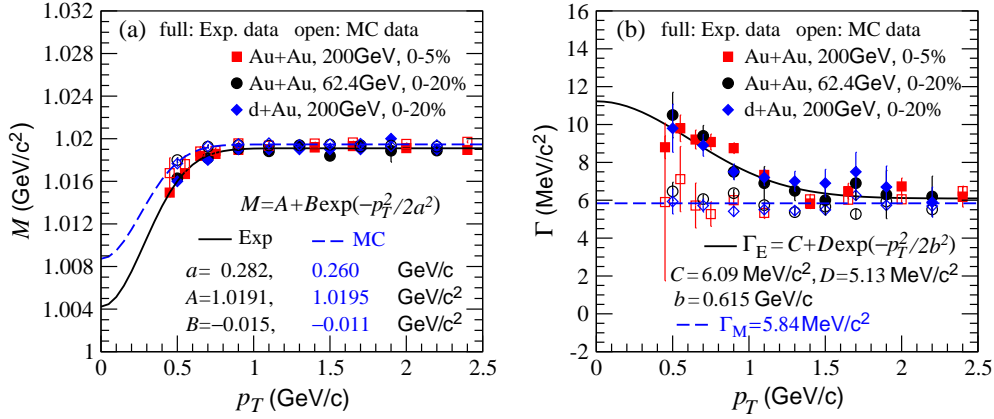}
\vspace*{-2mm}
\caption{Gaussian-formula fits to the experimental data and the Monte Carlo data
of the $K^+K^-$ mass and mass-width in the Au+Au collisions at $\sqrt{s_{NN}}=$200
and 62.4 GeV and the d+Au collisions at $\sqrt{s_{NN}}=$200 GeV \cite{STAR-PRC09}.}
\label{zf-mass-wid}
\end{figure}

In order to estimate the BBC in the collisions, we need to determine
the mass modification in the source medium.  In Ref.\ \cite{STAR-PRC09},
the $\phi$ mass and mass-width are investigated in the Au+Au and d+Au
collisions at the RHIC, by measuring the decayed $K^+K^-$ pair.  The results
show that there are more differences between the experimental data and the
Monte Carlo data in the low transverse momentum $p_T$ region \cite{STAR-PRC09}.
We show in Fig.\ \ref{zf-mass-wid} the Gaussian-formula fits to the experimental
data and the Monte Carlo data of the mass and mass-width \cite{STAR-PRC09} in 
the Au+Au and d+Au approximately central collisions at the RHIC.
The data of the three collisions have the similar variation with $p_T$.
At low $p_T$, the similar mass and mass-width modifications for the large
systems of the Au+Au collisions and the small system of the d+Au collision
imply that the $\phi$ mesons may have sufficient interactions with the medium
particles even for the small system, i.e., the $\phi$ mean free path is
smaller as compared to the small system size.  This is consistent with the
estimations of the $\phi$ mean free path between 1 and 2.4 fm in the hadronic
gas in the hidden local symmetry model \cite{AlvaKoch02}.
Phenomenologically, a particle with lower momentum moves more circuitously
in medium due to the collisions with the medium particles.  The $\phi$
mesons with lower average transverse momentum, thus with lower average
transverse velocity ${\bar v}_T$, have more possibility decaying inside
the source medium and the possibility will be higher if the source with
a transverse expanding velocity comparable to ${\bar v}_T$.  The differences
between the experimental data and the Monte Carlo data at low $p_T$ reflect
the medium effects on the measurements of the mass and mass-width.
On the other hand, the $\phi$ with higher momentum escapes the source more
easily.  The relativistic space shrink also makes the $\phi$ with high
relative velocity to the medium go through the medium easily.
In this case, the $\phi$ tends to decay outside of the source and the
medium effects on the mass and mass-width approach to zero.  However,
the BBC of the $\phi\phi$ pairs can remain.  In the reactions of 12 GeV
p on Cu targets, KEK-PS E325 Collaboration also observed the mass
modification of the $\phi$ in low velocity region \cite{E325-PRL07},
where the mediums are approximately rest for the lower energy collisions.

Assuming the measured invariant-mass distribution of $K^+K^-$ in the Au+Au
and d+Au collisions \cite{STAR-PRC09} consists of the two parts, one from
the contribution of the $\phi$ mesons decaying inside the source medium and
another from the contribution of the $\phi$ mesons decaying outside of the
source, we have the normalized density distribution of the mass as,
\begin{eqnarray}
&&\!\!\rho_{\rm exp}(m;M_{\rm exp},\Gamma_{\rm exp})
=\frac{\Gamma_{\rm exp}/2\pi}{(m-M_{\rm exp})^2+(\Gamma_{\rm exp}/2)^2}
~~~~~\nonumber\\
&&=f(p_T)\rho_0(m;M_0,\Gamma_0)\!+\![1\!-\!f(p_T)]\rho_*(m;M_*,\Gamma_*),~~
\label{nmdis}
\end{eqnarray}
where $M_{\rm exp}$ and $\Gamma_{\rm exp}$ are the mass center and width
measured in experiment, $M_0$ and $\Gamma_0$ are the mass center and width
for the $\phi$ mesons decaying outside of the source, and $M_*$ and $\Gamma_*$
are the mass center and width for the $\phi$ mesons decaying inside the source
medium.  In Eq.\ (\ref{nmdis}), $f$ is the fraction of the $\phi$ meson
decay outside of the source.  With the experimental data of ($M_{\rm exp}$,
$\Gamma_{\rm exp}$) \cite{STAR-PRC09} and with the corresponding Monte Carlo
data \cite{STAR-PRC09} instead of ($M_0$, $\Gamma_0$), we can obtain the
distributions $\rho_{\rm exp}(m)$ and $\rho_0(m)$, and then determine $M_*$,
$\Gamma_*$, and $f$ from Eq.\ (\ref{nmdis}).

In table \ref{Tab-Fit-f}, we present the results of $M_*$, $\Gamma_*$, and
$f$ obtained by fitting the distributions $\rho_{\rm exp}(m)$ with the formula
(\ref{nmdis}), where $\rho_{\rm exp}(m)$ and $\rho_0(m)$ are calculated with
the three experimental data and the Monte Carlo data of the mass and mass-width
shown in Fig.\ \ref{zf-mass-wid}.  For the Fit 1 in the table, we first perform
the fit in the lowest transverse momentum region with the average 0.5 GeV$/\!c$
and get the fitted results $M_*=1.0157$ GeV$/\!c^2$ and $\Gamma_*=9.785$
MeV$/\!c^2$.  Because the distribution $\rho_{\rm exp}(m)$ is calculated with
the data of ($M_{\rm exp}$, $\Gamma_{\rm exp}$) which have large differences
to the date of ($M_0$, $\Gamma_0$) in the momentum region, the fitted result
of $f$ is zero.  When the interaction between the medium particles and the
$\phi$ is sufficient, the effect of the source medium on the $\phi$ mass may
reach a saturation.  In this case, if we assume that the values of $M_*$ and
$\Gamma_*$ are momentum-independent, we can obtain the values of $f$ in the
higher $p_T$ regions as shown in the Fit 1 of table \ref{Tab-Fit-f}, by
performing the fits with the fixed $M_*$ and $\Gamma_*$ values.  The fitted
results of $f$ increase with $p_T$ when the differences between the experimental
data of ($M_{\rm exp}$, $\Gamma_{\rm exp}$) and the Monte Carlo data of ($M_0$,
$\Gamma_0$) decrease with $p_T$.

\begin{table*}[htb]
\caption{Results of fitting $\rho_{\rm exp}(m)$ with Eq.\ (\ref{nmdis}). }
\begin{tabular}{c|ccccc}
\hline\hline
\multicolumn{2}{c}{\hspace*{16mm}$p_T$(GeV$\!/\!c$)}&~~$~M_*$(GeV$\!/\!c^2$)&
~~~~~$\Gamma_*$(MeV$\!/\!c^2$)&~~$~~f$&~~$\chi^2$/NBF~\\
\hline
&~0.5~&~~$1.0157\pm0.0003$~&~~~$~9.785\pm0.373$~&~~~$~0.000\pm0.008$~~&~~0.04/30\\
~Fit 1~~&~0.7~&~~1.0157(fixed)~&~~~~9.785(fixed)~&~~~$~0.481\pm0.031$~~&~~65.05/30\\
&~1.0~&~~1.0157(fixed)~&~~~~9.785(fixed)~&~~~$~0.753\pm0.014$~~&~~68.43/30\\
\hline
&~0.5~&~~$1.0148\pm0.0005$~&~~~$~12.069\pm0.921$~&~~~0.40(fixed)~&~~0.56/30\\
~Fit 2~~&~0.7~&~~1.0148(fixed)~&~~~12.069(fixed)~&~~~$~0.640\pm0.021$~~&~~51.98/30\\
&~1.0~&~~1.0148(fixed)~&~~~12.069(fixed)~&~~~$~0.829\pm0.009$~~&~~65.82/30\\
\hline
&~0.5~&~~$1.0107\pm0.0016$~&~~~$~20.197\pm5.773$~&~~~0.80(fixed)~&~~5.95/30\\
~Fit 3~~&~0.7~&~~1.0107(fixed)~&~~~20.197(fixed)~&~~~$~0.846\pm0.009$~~&~~71.80/30\\
&~1.0~&~~1.0107(fixed)~&~~~20.197(fixed)~&~~~$~0.926\pm0.004$~~&~~111.79/30\\
\hline\hline
\end{tabular}
\label{Tab-Fit-f}
\end{table*}

In the fit with Eq.\ (\ref{nmdis}), the fitted results of $\Gamma_*$ and $f$
are highly correlated when $f$ is not zero.  In this case it is hard to get
the appropriate results of them meanwhile in the fit.
If we assume that only 60 percent and even 20 percent of the $\phi$ mesons
in the lowest transverse momentum region decay inside the source medium, we
can obtain the fitted results ($M_*=1.0148$ GeV$/\!c^2$, $\Gamma_*=12.069$
MeV$/\!c^2$) presented in the Fit 2 of table \ref{Tab-Fit-f} by fixing $f=0.4$
and the fitted results ($M_*=1.0107$ GeV$/\!c^2$, $\Gamma_*=20.197$ MeV$/\!c^2$)
presented in the Fit 3 of table \ref{Tab-Fit-f} by fixing $f=0.8$, respectively.
With the same way, we can also obtain the fitted results of $f$ in the momentum
regions of higher $p_T$.  The fitted results of $f$ in the Fit 2 and Fit 3 of
table \ref{Tab-Fit-f} are larger than those in the Fit 1 of table \ref{Tab-Fit-f}.
In the fits, the increase of $f$ with increasing $p_T$ implies more and more the
$\phi$ mesons decaying outside of the source medium with increasing $p_T$.
Considering that not all the $\phi$ mesons in the transverse momentum region
of $p_T=0.5$ GeV/$c$ decay inside the source medium, the values of $M_*$ and
$\Gamma_*$ will have larger differences to the values of $M_0$ and $\Gamma_0$,
respectively, compared to those for the Fit 1 of table \ref{Tab-Fit-f}.  This
will lead to stronger BBC.

\begin{figure}[!htbp]
\vspace*{2mm}
\includegraphics[scale=0.75]{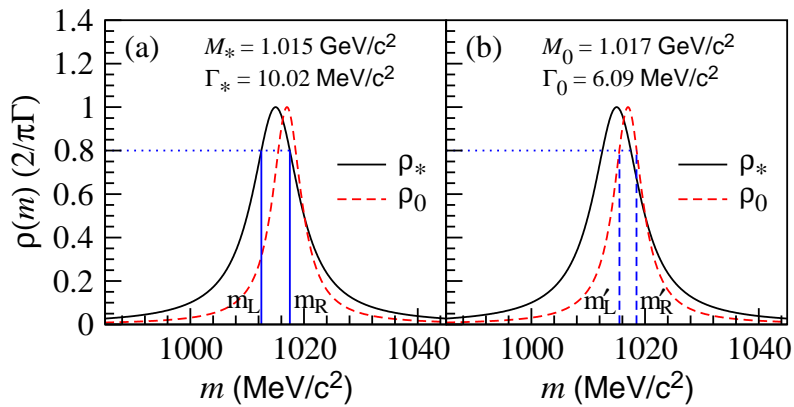}
\vspace*{-2mm}
\caption{Mass distributions $\rho_*$ and $\rho_0$ for the Au+Au collisions
at $\sqrt{s_{NN}}=200$ GeV. }
\label{zfrho-LR}
\end{figure}

Once obtaining the values of $M_*$ and $\Gamma_*$, we can calculate the BBC
functions with the hydrodynamic code for the heavy-ion collisions.
As shown in Fig.\ \ref{zfrho-LR} for the Au+Au collisions at $\sqrt{s_{NN}}
=$200 GeV for instance, the mass $m_L$($m_R$) in Fig.\ \ref{zfrho-LR}(a),
which is at the left(right) of the peak, is taken with relative probability
$\frac{1}{2}0.8$ in the calculations for the mass in the source medium, and
the corresponding mass out of the source is $m'_L$($m'_R$) shown in Fig.
\ref{zfrho-LR}(b).  Here, the values of $M_*$ and $\Gamma_*$ are taken as in
the Fit 1 in table \ref{Tab-Fit-f}, $M_0$ is taken to be the Gaussian-formula
fit result for the corresponding Monte Carlo data, and $\Gamma_0$ is taken to
be 6.09 MeV$\!/c^2$ (see Fig. \ref{zf-mass-wid}).

\begin{figure}[!htbp]
\vspace*{4mm}
\includegraphics[scale=0.58]{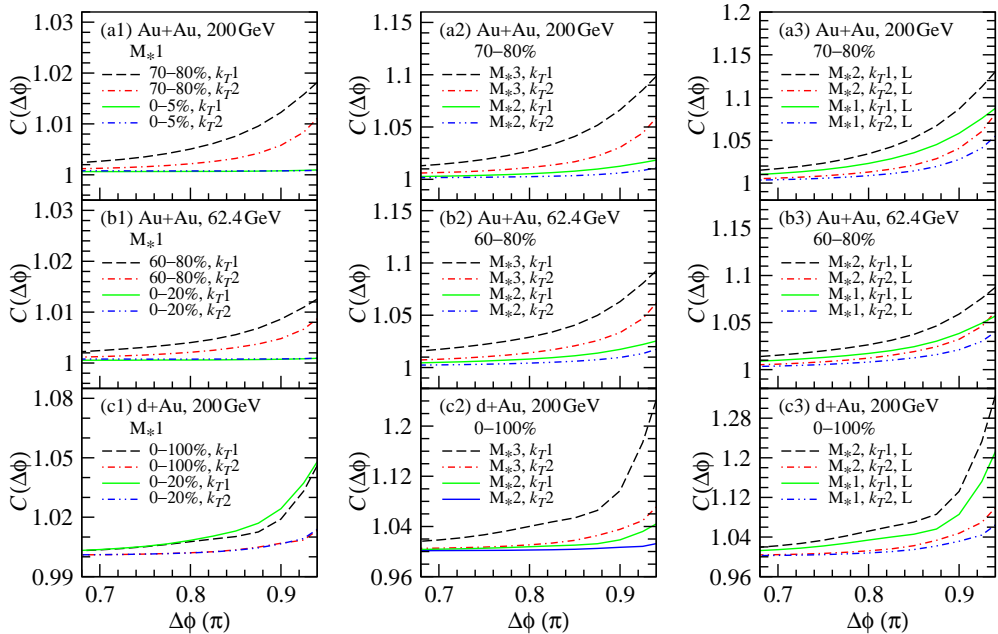}
\vspace*{-4mm}
\caption{BBC functions of $\phi\phi$ for the collisions of Au+Au and d+Au
at the RHIC energies. }
\label{BBC-Dphi}
\end{figure}

We show in Fig.\ \ref{BBC-Dphi} the BBC functions $C(\Delta\phi)$ for the
collisions of Au+Au and d+Au at the RHIC energies.  Here, $\Delta\phi$ is
the angle between the transverse momenta $\mk_{1T}$ and $\mk_{2T}$ of the
two $\phi$ mesons, and the results for the marks of $k_T1$ and $k_T2$ are
calculated in the higher and lower momentum regions $0.4<|\mk_T|<0.5$
GeV/$c$ and $0.8<|\mk_T|<0.9$ GeV/$c$, respectively.  We take 1000 events
for each of the collisions.  The masses of the $\phi$ mesons in and out the
source medium are taken with the distributions $\rho_*(m)$ and $\rho_0(m)$
as explained with Fig.\ \ref{zfrho-LR}.  In Figs.\ \ref{BBC-Dphi}(a1)--(c1),
label ``M$_*1$" means that the values of $M_*$ and $\Gamma_*$ for calculating
$\rho_*(m)$ are taken as the fit results in the Fit 1 of table \ref{Tab-Fit-f}.
One can see that the $C(\Delta\phi)$ results for the peripheral Au+Au
collisions are larger than those for the central collisions, because
the peripheral collisions have narrower source temporal distributions
\cite{Zhang15}.  For the same reason, the results for d+Au collisions
are large.  For the asymmetric collisions of d+Au, the BBC function
$C(\Delta\phi)$ for 0--100\% centrality is slight narrower than that
for 0--20\% centrality.  The main reason is that the more event-by-event
fluctuations of the BBC function \cite{Zhang15a} in the collisions of a
wider centrality range lead to smaller averaged results \cite{Zhang15}.
In Figs. \ref{BBC-Dphi}(a2)--(c2), the in-medium mass distributions
$\rho_*(m)$ are calculated with the $M_*$ and $\Gamma_*$ which are taken
as the fit results in the Fit 2 of table \ref{Tab-Fit-f} (labeled M$_*2$)
and in the Fit 3 of table \ref{Tab-Fit-f} (labeled M$_*3$), respectively.
In these cases, the BBC functions are larger than those in the M$_*1$ case,
which are the lower limits.  In Figs. \ref{BBC-Dphi}(a3)--(c3), the results
labeled ``left" are calculated with only the $\phi$ mesons which with the
masses smaller than the maximum (at left of the peak).  The interactions
in medium lead to the decrease of $M_*$ to $M_0$ and the increase of
$\Gamma_*$ to $\Gamma_0$.  This makes the ``left" particles have larger
mass shifts (see Fig.\ \ref{zfrho-LR}), and thus have stronger BBC.

In summary, we have presented an analysis of the medium effect on the $\phi$
mass in the Au+Au and d+Au collisions at the RHIC and investigated the BBC
of $\phi\phi$ pairs in the collisions.  We find that the invariant mass and
mass-width of $K^+K^-$ measured in the collisions are consistent with a picture
in which the $\phi$ with low transverse momentum has more possibility decaying
inside the source medium than the $\phi$ with high transverse momentum.
The investigations indicate that the BBC may perhaps be observed in the
collisions of d+Au and the peripheral collisions of Au+Au at the RHIC.
We suggest to measure the BBC experimentally for understanding the mass
modifications of the $\phi$ meson in the collisions.

\vspace*{-4mm}
\begin{acknowledgments}
\vspace*{-4mm}
We thank J. H. Chen and G. Wang for helpful discussions of the experimental
and Mote Carlo data of $\phi$ meson.  We thank C. Shen and H. Song for helpful
discussions of the viscous hydrodynamic code.
This research was supported by the National Natural Science Foundation of
China under Grant Nos. 11675034, 11275037, and 11647166.
\end{acknowledgments}

\end{document}